\documentclass{sf2a-conf2023}
\usepackage{graphicx}
\usepackage{hyperref}
\usepackage[]{natbib}  
\usepackage{epstopdf}

\def\BibTeX{{\rm B\kern-.05em{\sc i\kern-.025em b}\kern-.08em
    T\kern-.1667em\lower.7ex\hbox{E}\kern-.125emX}}
\bibpunct{(}{)}{;}{a}{}{,}  

\begin{document}

\TitreGlobal{SF2A 2023}

\title{Challenges and limitations of future exoplanet space imagers}

\runningtitle{Short title here}

\author{L. Leboulleux}\address{Univ. Grenoble Alpes, CNRS, IPAG, 38000 Grenoble, France}

\setcounter{page}{237}


\maketitle


\begin{abstract}
The James Webb Space Telescope was not even launched yet when the Astro2020 Decadal Survey American report recommended the development of what is now called the Habitable World Observatory, also mentioned by the Voyage 2050 European report. This future space telescope, at $11$ billions dollars and at least $6$ m diameter, should allow, around $2040$, the characterization of at least 25 exoplanets similar to Earth and orbiting around main sequence stars, with the hope of discovering one where life could have developed. This objective represents a technological challenge since it requires the design of spectro-imagers able to access very high contrasts ($10^{-8}-10^{-10}$) at low angular separations (smaller than $100$ mas). This proceeding and the talk it is associated to address various obstacles that remain to be overcome in order to one day allow HabWorld to reach its ultimate performance.
\end{abstract}



\section{Introduction}
Every ten years, the National Academies of Sciences, Engineering, and Medicine (NASEM) publish recommendations for large science projects in various fields. In space astronomy for instance, the Chandra-X ray Observatory and the Hubble Space Telescope followed the "Astronomy and Astrophysics for the 1980s" NASEM report, the Spitzer Space Telescope issued from the 1991 "The Decade of Discovery in Astronomy and Astrophysics" report, and more recently, the Next Generation Space Telescope, now known as the James Webb Space Telescope (JWST), was recommended by the 2001 "Astronomy and Astrophysics in the New Millennium" report \citep{Gardner2023}. In 2010, the "New Worlds, New Horizons" NASEM report recommended to push forward the research in dark energy with the so-called Nancy Grace Roman Space Telescope (RST), a telescope that will be launched around $2028$. Its science goal was extended to the field of exoplanet imaging and characterization, with the first detection of exoplanets in reflected light. It is now also equiped with a coronagraphic instrument able to detect and characterize planets down to $200-300$ mas from their host star and with contrasts down to $10^{-8}-10^{-9}$ ($R \sim 50$ at $700$ nm) \citep{Bailey2023}. In 2021, a new decadal report was released: the "Pathways to Discovery in Astronomy and Astrophysics for the 2020s" recommended the development of the so-called Habitable World Observatory (HabWorlds). This new large space mission should be able to characterize $25$ Earth-like planets with separations of $20-50$ mas, contrasts down to $10^{-10}$ at mid to high spectral resolution ($200-5000$ at $500$ nm) \citep{Vaughan2023, Roberge2021, Gaudi2021, Gaudi2020}. 

With these two future space missions, the scientific requirements reach values that have never been accessed before, from space and from the ground. As a comparison, they imply a gain in contrast of a factor $10^{5}$, and a gain in angular separation of a factor $20$ compared to the JWST constraints. In addition to the difficulty to access such extreme performance in perfect conditions, it is highly deteriorated as soon as conditions are non perfect. As an example, one of the key phases in the JWST launch was the segment phasing of its primary mirror: $18$ hexagonal segments had to be aligned from a few microns to $30$ nm RMS of aberration, a complex operation that took one month to be completed. If HabWorlds, whose primary mirror will probably also be segmented, is aligned down to $30$ nm RMS, its contrast will be maintained over $10^{-6}$, far from its requirements. Compared to the JWST or other telescopes, the technological breakthrough must concern both the absolute performance and its robustness to realistic conditions.

This proceeding is dedicated to describing methods and tools that are developed or under development to access high contrasts at small angular separations (section \ref{s:Performance}) and, despite non perfect conditions (section \ref{s:Limitations}), maintain such a performance over time and observation conditions (section \ref{s:Robustness}).

\section{Performance: coronagraphy and wavefront control for (non) dummies}
\label{s:Performance}

\begin{figure}[ht!]
 \centering
 \includegraphics[width=0.53\textwidth,clip]{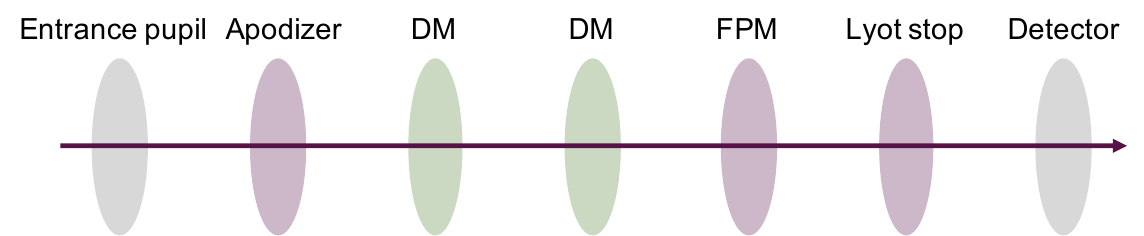}
  \caption{Scheme of a typical high-contrast imaging system. Purple: components associated with coronagraphy, Green: components associated with wavefront control. DM stands for Deformable Mirror and FPM for Focal Plane Mask.}
  \label{Leboulleux:fig1}
\end{figure}

The access to high contrasts in exoplanet imaging or spectroscopy requires the assembly of multiple components, set up in pupil planes, focal planes, or out of pupil planes. The general principle is illustrated in Fig.~\ref{Leboulleux:fig1} and is made of a coronagraph, which can be combined with a wavefront control system, with the ultimate objective of removing the starlight in the area of interest of the detector plane.

The coronagraph is made of up to three passive components :

$\bullet$ An apodizer (pupil plane) that reshapes the wavefront to modulates its intensity in the focal plane,

$\bullet$ A Focal-Plane Mask (FPM) that blocks the starlight along the optical axis,

$\bullet$ A Lyot stop (pupil plane) to remove the high-order star signal, not stopped by the FPM.

\begin{figure}[ht!]
 \centering
 \includegraphics[width=0.47\textwidth,clip]{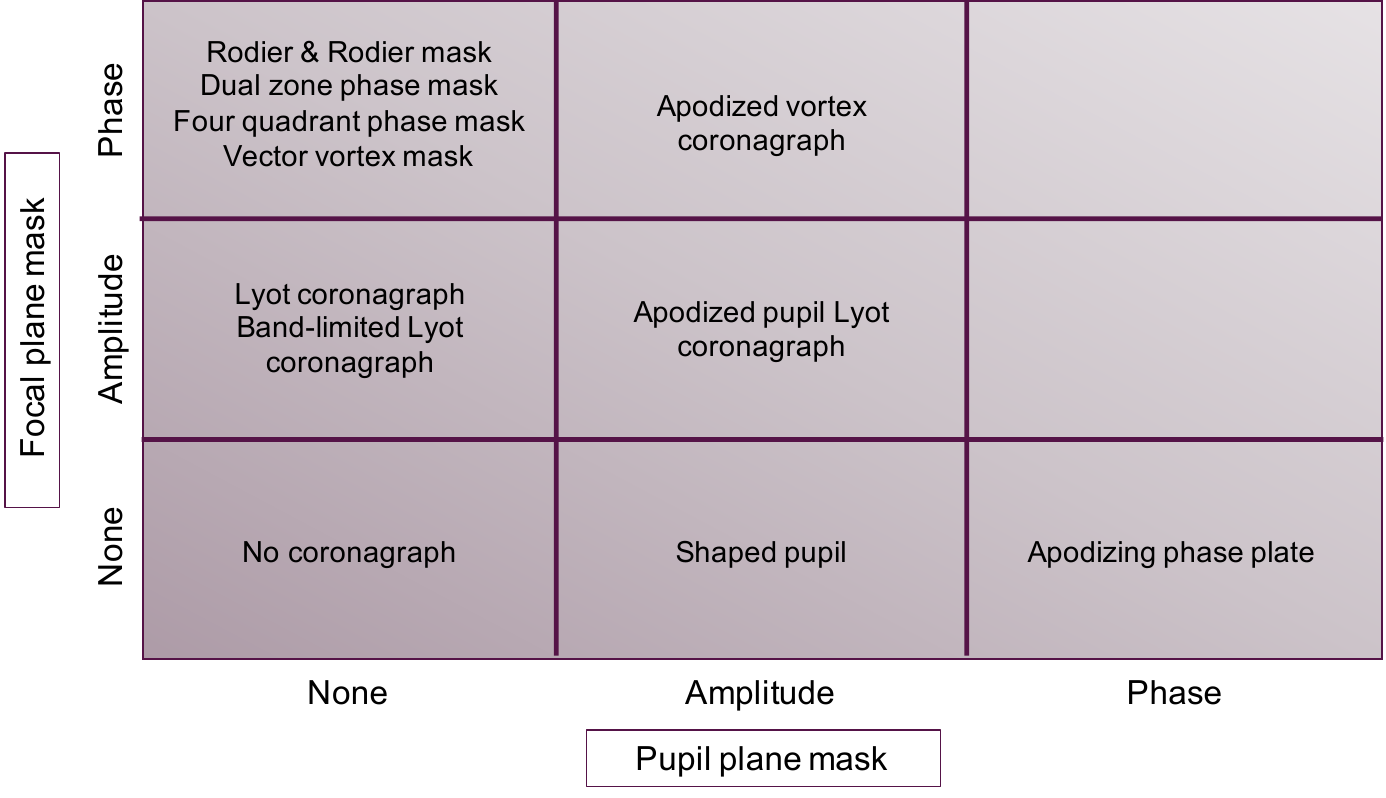} \includegraphics[width=0.52\textwidth,clip]{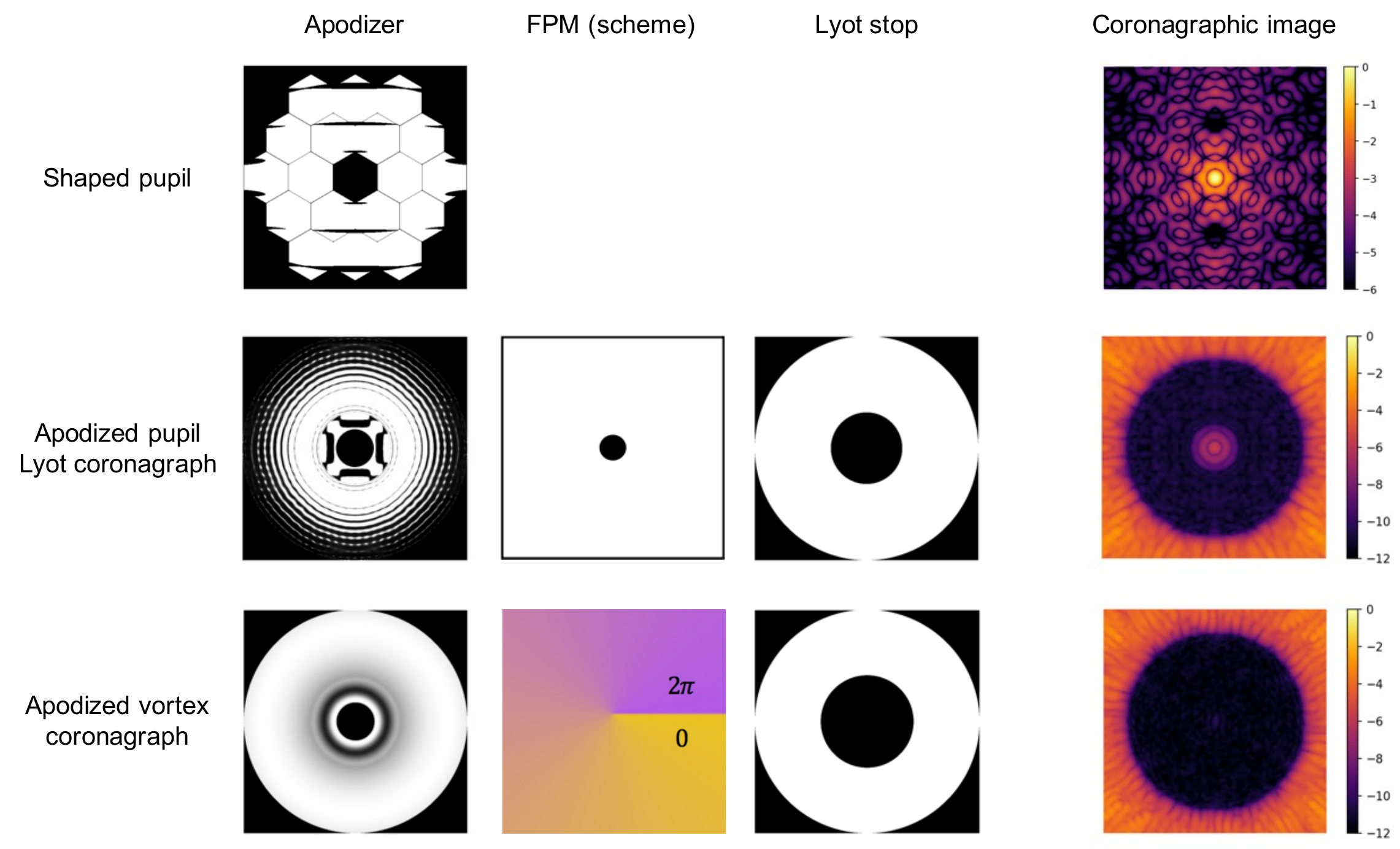}
  \caption{(left) Non exhaustive list of coronagraphs classified by their types of focal plane and pupil plane masks. (right) Examples of coronagraphs and their associated coronagraphic images (no aberration added). The shaped pupil is taken from \cite{Leboulleux2022} and the apodized pupil Lyot coronagraph and the apodized vortex coronagraph from \cite{Mazoyer2018, N'Diaye2015, Fogarty2017}. The focal-plane images are in logarithmic scales.}
  \label{Leboulleux:fig2}
\end{figure}

The apodizer and the FPM, if they exist, can be amplitude or phase masks, and a large variety of combinations are possible depending on the type of masks. Fig.~\ref{Leboulleux:fig2} (left) introduces a few of these combinations, with no mention of the existence of a Lyot stop. For a more exhaustive list, please refer to \cite{Ruane2018}. Fig.~\ref{Leboulleux:fig2} (right) also illustrates this diversity with three cases: 1) a shaped pupil (amplitude apodizer, no FPM, no Lyot stop, see also \cite{Kasdin2003, Carlotti2011}) optimized for a 18 segment pupil to dig a circular $1 \lambda/D$-large dark zone localized at $8.3 \lambda/D$, 2) an apodized pupil Lyot coronagraph (amplitude apodizer, FPM, and Lyot stop, see also \cite{Aime2002, Soummer2006, N'Diaye2016}) to access high contrasts from $3$ to $10 \lambda/D$, and 3) an apodized vortex coronagraph (amplitude apodizer, phase FPM, and amplitude Lyot stop, see also \cite{Carlotti2013, Fogarty2017}) to access high contrasts from $3$ to $10 \lambda/D$.

In addition of a coronagraph, a wavefront control system enables to reach higher contrasts at small angular separations by reshaping the wavefront with one or two Deformable Mirrors (DM). One DM is set in a pupil plane for phase control, but with one single DM, only half of the detector interest zone can be controled. A second DM can be installed out of pupil plane to access amplitude control and a symetric dark area. Due to the limited number of actuators, phase and amplitude are controlled with a small spatial sampling, which makes this tool usable only for small dark zones at close angular separations: with a $34$-actuators across the diameter DM, the dark hole cannot go further than $17 \lambda/D$. More details, with the descriptions and references of the diverse methods, can be found in \cite{Jovanovic2018} (see also figure 4 for illustration).

\section{Limitations}
\label{s:Limitations}

\begin{figure}[ht!]
 \centering
 \includegraphics[width=0.77\textwidth,clip]{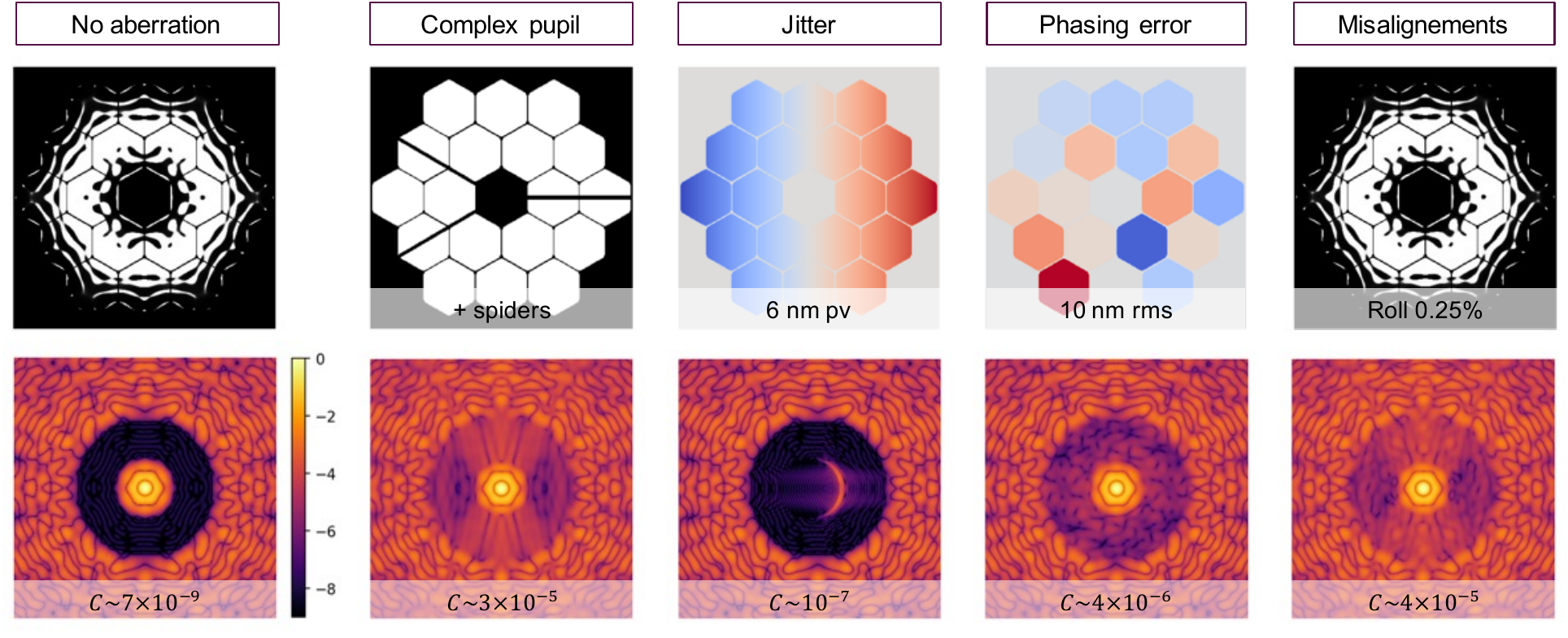}
  \caption{Examples of aberrations and their impact on the coronagraphic image with a 18-segment pupil combined with a shaped pupil. From left to right: no aberration, spiders added on the pupil, tip or jitter (a FPM has been added in the coronagraphic system), segment phasing errors, apodizer vertical shift.}
  \label{Leboulleux:fig4}
\end{figure}

Coronagraphs, no matter their ultimate performance, are highly sensitive to unperfect conditions. Fig.~\ref{Leboulleux:fig4} illustrates a few of them for space operations, in the case of a shaped pupil compatible with a $18$ segment pupil designed to access an ultimate contrast of $7 \times 10^{-9}$ between $6$ and $13 \lambda/D$:

$\bullet$ When the pupil is complexified, for instance here with spiders added on the architecture: this is the reason why HabWorlds is considered off-axis, to avoid a complex pupil with spider and central obstruction. 

$\bullet$ With global tip-tilt misalignments (or pointing error or jitter): they are particularly problematic in case of a FPM since the starlight leaks off the FPM at small angular separations. 

$\bullet$ With phasing errors: on HabWorlds, a contrast of $10^{-10}$ imposes the phasing errors to remain below $10$pm rms, which is not accessible yet in labs. 

$\bullet$ With a mask shift or misalignment.

The high-contrast and high-angular separation performance can then not be accessed without a robust aberration mitigation. This mitigation can take two forms:

$\bullet$ Cure: diagnosing the sources of errors (wavefront sensing or WFS) and correcting them (wavefront sensing or WFC) with adaptive components,

$\bullet$ Prevention: optimizing the instrument to make it insensitive to errors. 

\section{Robustness: prevention or cure?}
\label{s:Robustness}

\subsection{Cure}

Curing relies on two steps:

$\bullet$ a diagnosis of the aberrations or WFS. Reconstructing the errors, static or dynamic, is complex and several methodologies and tools have been developed so far. They are not described in the proceeding and for more information please refer to \cite{Jovanovic2018}.

$\bullet$ a correction of the aberrations or WFC. Several methods have also been developed, all of them requiring the use of adaptive components \citep{Snik2018}.

As phase control tools, DMs \citep{Blanco2022, Bowens-Rubin2021} have been used since the beginning of adaptive optics \citep{Chauvin2018} and can enable to compensate for the aberrations and recover the performance. One of its latest application is the combination of two DMs and its use through the Active Correction of Aperture Discontinuities to compensate for both aperture complexity (spiders) and phase errors (segment phasing misalignements) on a space coronagraphic system \citep{Mazoyer2018, Mazoyer2018b}. If DMs have so far never been used in space (except on specific experiments on balloon missions, see \cite{Morgan2019}), the Nancy Grace Roman Space Telescope will be the first space telescope equipped of two deformable mirrors \citep{Mennesson2022}. As an alternative for phase control, one can cite the spatial-light modulator (SLM) technology \citep{Kühn2022, Kühn2018} which can be used, between other applications, as a continuous and low- to high-order programmable pupil phase mirror, despite being limited by its requirement to reflect linear polarized light. More recently, investigations have been ongoing to identify adaptive amplitude components. The use of Digital Micro-mirror Devices (DMD) is currently under study: the component tested at IPAG, for instance, is built as a matrix of small square mirrors that can be controlled in tip-tilt and flipped between two positions, one of them used as a transmissive direction towards the detector, the other one rejecting the light out of the optical system. It can then be used to get programmable binary apodizations \citep{Carlotti2023, Leboulleux2022a, Carlotti2018}. 

\begin{figure}[ht!]
 \centering
 \includegraphics[width=0.72\textwidth,clip]{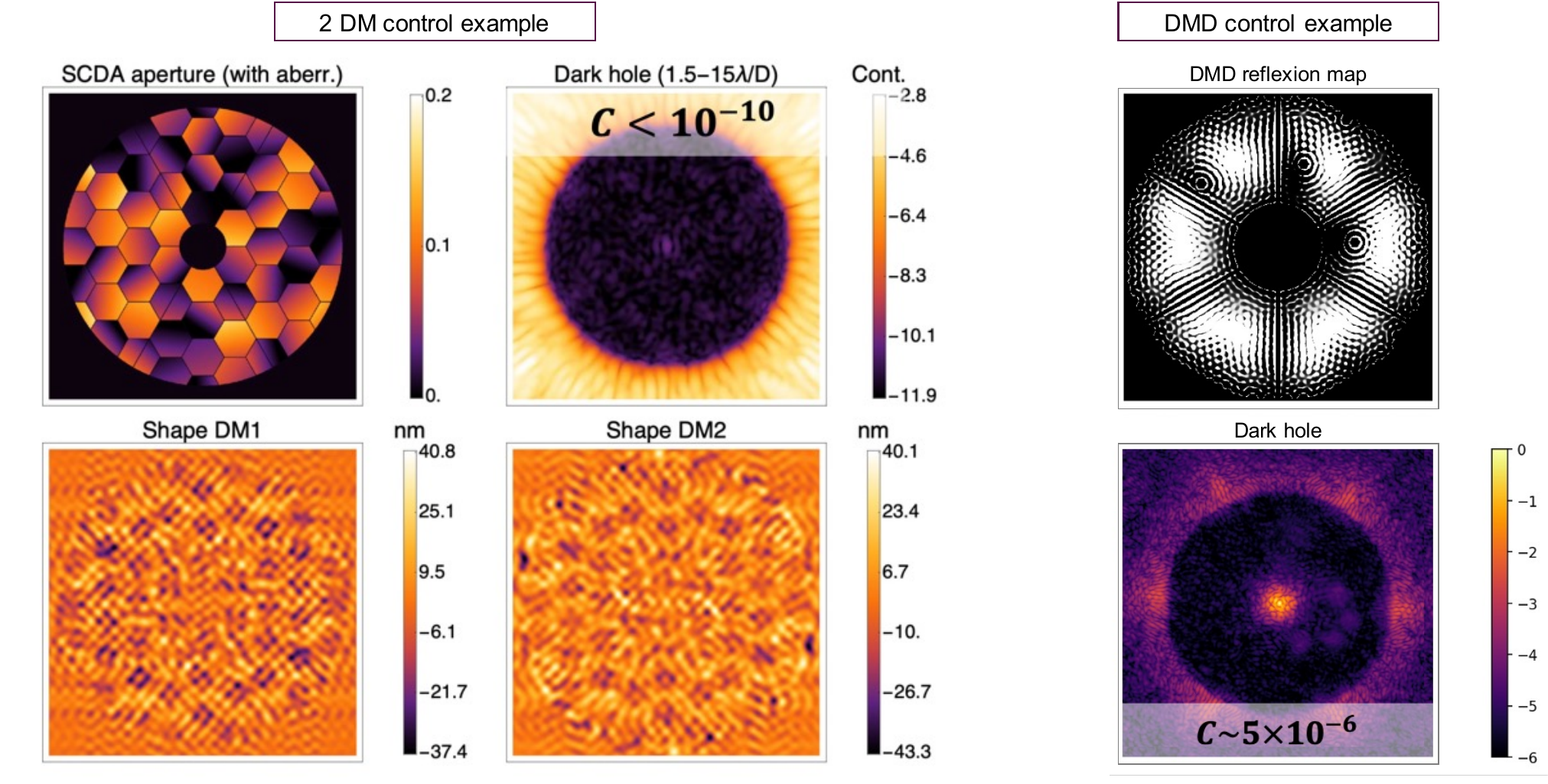}
  \caption{Examples of aberration mitigation with adaptive components: 1) use of two deformable mirrors to recover the dark hole despite segment phasing errors with the ACAD-OSM methodology (simulation from \cite{Mazoyer2018}), 2) use of a DMD to recover the dark hole despite missing segments (experimental image from \cite{Carlotti2023}).}
  \label{Leboulleux:fig5}
\end{figure}

Fig.~\ref{Leboulleux:fig5} illustrates the use of 1) two DMs to mitigate the impact of phasing errors in the coronagraph dark region, 2) one DMD to recover the contrast in the coronagraph dark region despite missing segments.

\subsection{Prevention}

An alternative approach is to create a high-contrast system that is passively blind to or robust to aberrations by design, meaning optimized to remain performant (high contrast at small angular separations) despite a range of aberrations. As aberrations have a high impact on the performance, the robustness is a key aspect of the design study. Different methods have been proposed so far to increase the system robustness, combined with different coronagraphs: to reduce the impact of low-order aberrations on vortex coronagraphs \citep{Fogarty2020}, scalar vortex coronagraphs \citep{Desai2022}, or apodized pupil Lyot coronagraphs \citep{N'Diaye2015}, or Lyot stop shifts on apodized pupil Lyot coronagraphs \citep{Nickson2022}.

\begin{figure}[ht!]
 \centering
 \includegraphics[width=0.69\textwidth,clip]{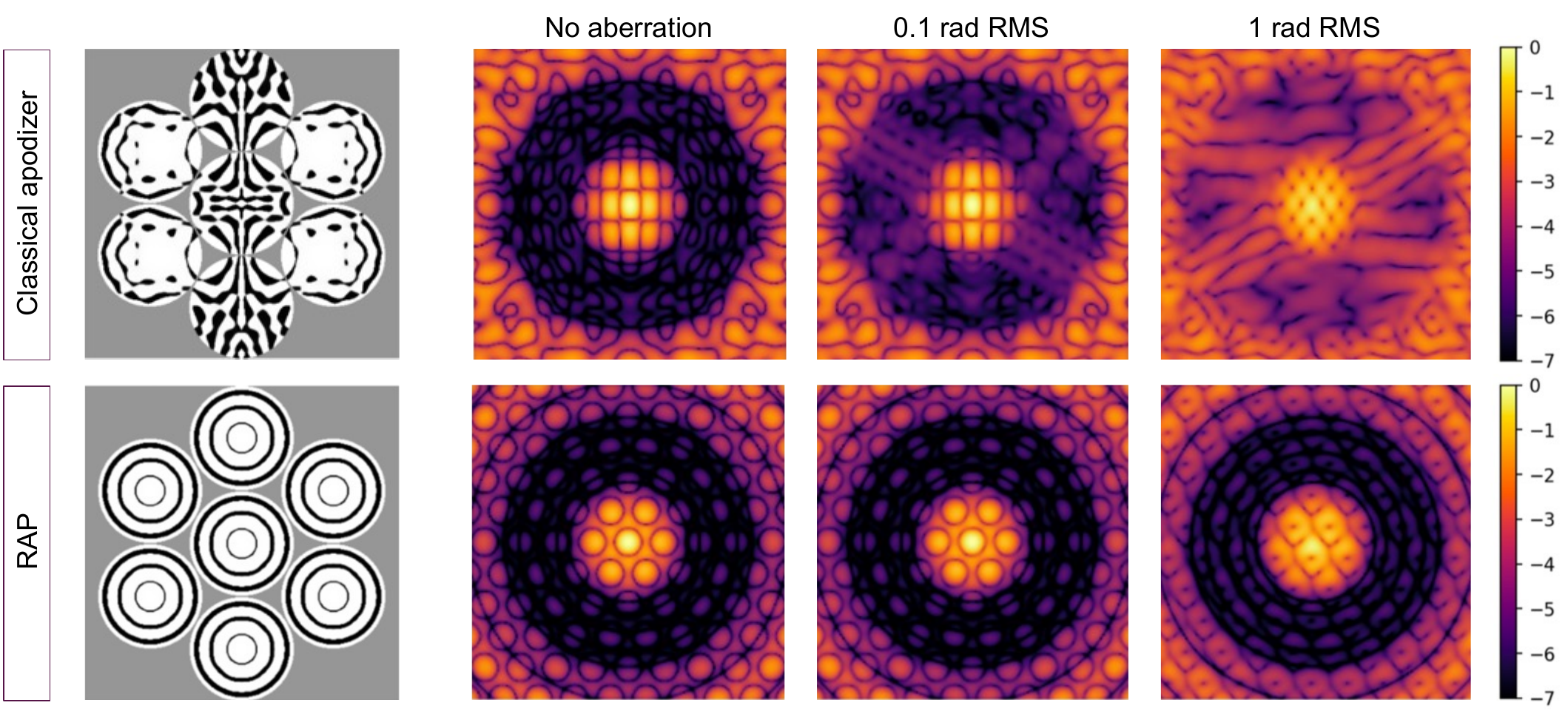}
  \caption{Robustness of RAP design for segment-phasing errors on a Giant Magellan Telescope-like pupil (7 circular segments): (left) two apodizing phase plate maps, with $\pi$-shift between black and white areas, the top apodizer is classically optimized, the bottom one is redundantly optimized; (right) associated coronagraphic PSFs with increasing segment phasing errors (figure from \cite{Leboulleux2022c}).}
  \label{Leboulleux:fig6}
\end{figure}

Let's focus on one example only, with a fully analytical construction: \cite{Leboulleux2018, Laginja2021} introduce the Pair-based Analytical model for Segmented Telescopes Imaging from Space (PASTIS) to evaluate the impact of phasing errors on the coronagraphic PSF and performance. In the PASTIS formula, a specific factor appears, as a weight factor of all other factors: the PSF of one segment only. If this weight factor, so the PSF of the segment, increases, then the contrast is easily deteriorated by segment-level aberrations. On the opposite, if the PSF of the segment is minimal, then the contrast is less impacted by any segment-level aberration. This assessment led to the Redundant Apodized Pupil (RAP) concept \cite{Leboulleux2022c, Leboulleux2022b}: optimizing the segment and multiplying this apodization to mimic the pupil segmentation enable to obtain a high-contrast region that remains stable over a large range of segment-level aberrations, like segment phasing errors. Fig.~\ref{Leboulleux:fig6} illustrates this concept and its robustness for a $10^{-6}$-contrast dark zone between $5.5$ and $12 \lambda/D$.

\section{Conclusions}

The field of high-contrast imaging is the environment of many ongoing studies, that could not be all covered in this proceeding. The list of scientific requirements, with contrast and angular separations, is here far from exhaustive, since the planet photon flux, spectroscopy, polarimetry, or compatibility to post-processing methods were not even mentioned so far. Even with these requirements left on the side, the future scientific and technological ambitions for the Nancy Grace Roman Space Telescope and the Habitable Worlds Observatory demand multiple and complex instrumentation developments and research. Between others \citep{Laginja2022, Mazoyer2019}, in France, one can mention the development of several highly performant high-contrast testbeds, such as the Très Haute Dynamique one (THD2, \cite{Potier2019}), the Segmented Pupil Experiment for Exoplanet Detection (SPEED, \cite{Martinez2023}), the Marseille Imaging Testbed for High Contrast (MITHiC, \cite{ElMorsy2022}), or the High Contrast Module testbed (HCM, \cite{Hours2022}). 

These developments are necessary to overcome the limitations of current instruments and access rocky exoplanets. In space, the most efficient coronagraphs can currently not go deeper than contrasts of $\sim 10^{-5}$ at several hundreds of mas, and DMs have never been tested in space telescope, except for balloon missions. The Nancy Grace Roman Space Telescope and the Habitable Worlds Observatory are then setting ambitious objectives both in terms of science targets but also for first technological demonstrations.

\begin{acknowledgements}
The author thanks the SF2A organization committees (SOC and LOC) who made this conference a large success. She is also grateful for the help and references provided by colleagues, including Marie Ygouf, Elodie Choquet, Johan Mazoyer, Mamadou N'Diaye, Pierre Baudoz, and Alexis Carlotti.
\end{acknowledgements}

\bibliographystyle{aa}
\bibliography{Leboulleux1}

\begin{thebibliography}{43}
\expandafter\ifx\csname natexlab\endcsname\relax\def\natexlab#1{#1}\fi

\bibitem[{{Aime} {et~al.}(2002){Aime}, {Soummer}, \& {Ferrari}}]{Aime2002}
{Aime}, C., {Soummer}, R., \& {Ferrari}, A. 2002, \aap, 389, 334

\bibitem[{{Bailey} {et~al.}(2023){Bailey}, {Bendek}, {Monacelli}, {Baker},
  {Bedrosian}, {Cady}, {Douglas}, {Groff}, {Hildebrandt}, {Kasdin}, {Krist},
  {Macintosh}, {Mennesson}, {Morrissey}, {Poberezhskiy}, {Subedi}, {Rhodes},
  {Roberge}, {Ygouf}, {Zellem}, {Zhao}, \& {Zimmerman}}]{Bailey2023}
{Bailey}, V.~P., {Bendek}, E., {Monacelli}, B., {et~al.} 2023, arXiv e-prints,
  arXiv:2309.08672

\bibitem[{{Blanco} {et~al.}(2022){Blanco}, {Behara}, {Valenzuela}, \&
  {Kasper}}]{Blanco2022}
{Blanco}, L., {Behara}, N., {Valenzuela}, J., \& {Kasper}, M. 2022, in Society
  of Photo-Optical Instrumentation Engineers (SPIE) Conference Series, Vol.
  12185, Adaptive Optics Systems VIII, ed. L.~{Schreiber}, D.~{Schmidt}, \&
  E.~{Vernet}, 121856B

\bibitem[{{Bowens-Rubin} {et~al.}(2021){Bowens-Rubin}, {Hinz}, {Kuiper}, \&
  {Dillon}}]{Bowens-Rubin2021}
{Bowens-Rubin}, R., {Hinz}, P., {Kuiper}, S., \& {Dillon}, D. 2021, in Society
  of Photo-Optical Instrumentation Engineers (SPIE) Conference Series, Vol.
  11823, Techniques and Instrumentation for Detection of Exoplanets X, ed.
  S.~B. {Shaklan} \& G.~J. {Ruane}, 118231R

\bibitem[{{Carlotti}(2013)}]{Carlotti2013}
{Carlotti}, A. 2013, \aap, 551, A10

\bibitem[{{Carlotti} {et~al.}(2023){Carlotti}, {Baccar}, {Leboulleux},
  {Curaba}, {Delboulbé}, {Jocou}, {Moulin}, {Gluck}, \&
  {Sztefek}}]{Carlotti2023}
{Carlotti}, A., {Baccar}, S., {Leboulleux}, L., {et~al.} 2023, in Adaptive
  Optics for Extremely Large Telescopes VII (AO4ELT7)

\bibitem[{{Carlotti} {et~al.}(2018){Carlotti}, {Mouillet}, {Correia}, {Jocou},
  {Bourdarot}, {le Coarer}, {Forveille}, {Bonfils}, \& {Moulin}}]{Carlotti2018}
{Carlotti}, A., {Mouillet}, D., {Correia}, J.-J., {et~al.} 2018, in Society of
  Photo-Optical Instrumentation Engineers (SPIE) Conference Series, Vol. 10706,
  Advances in Optical and Mechanical Technologies for Telescopes and
  Instrumentation III, ed. R.~{Navarro} \& R.~{Geyl}, 107062M

\bibitem[{{Carlotti} {et~al.}(2011){Carlotti}, {Vanderbei}, \&
  {Kasdin}}]{Carlotti2011}
{Carlotti}, A., {Vanderbei}, R., \& {Kasdin}, N.~J. 2011, Optics Express, 19,
  26796

\bibitem[{{Chauvin}(2018)}]{Chauvin2018}
{Chauvin}, G. 2018, in Society of Photo-Optical Instrumentation Engineers
  (SPIE) Conference Series, Vol. 10703, Adaptive Optics Systems VI, ed. L.~M.
  {Close}, L.~{Schreiber}, \& D.~{Schmidt}, 1070305

\bibitem[{{Desai} {et~al.}(2022){Desai}, {Llop-Sayson}, {Bertrou-Cantou},
  {Ruane}, {Eldorado Riggs}, {Serabyn}, \& {Mawet}}]{Desai2022}
{Desai}, N., {Llop-Sayson}, J., {Bertrou-Cantou}, A., {et~al.} 2022, in Society
  of Photo-Optical Instrumentation Engineers (SPIE) Conference Series, Vol.
  12180, Space Telescopes and Instrumentation 2022: Optical, Infrared, and
  Millimeter Wave, ed. L.~E. {Coyle}, S.~{Matsuura}, \& M.~D. {Perrin}, 121805H

\bibitem[{{El Morsy} {et~al.}(2022){El Morsy}, {Vigan}, {Lopez}, {Otten},
  {Choquet}, {Madec}, {Costille}, {Sauvage}, {Dohlen}, {Muslimov}, {Pourcelot},
  {Floriot}, {Benedetti}, {Blanchard}, {Balard}, \& {Murray}}]{ElMorsy2022}
{El Morsy}, M., {Vigan}, A., {Lopez}, M., {et~al.} 2022, \aap, 667, A171

\bibitem[{{Fogarty} {et~al.}(2020){Fogarty}, {Mawet}, {Mazoyer}, {Sirbu},
  {Ruane}, \& {Pueyo}}]{Fogarty2020}
{Fogarty}, K., {Mawet}, D., {Mazoyer}, J., {et~al.} 2020, in Society of
  Photo-Optical Instrumentation Engineers (SPIE) Conference Series, Vol. 11443,
  Society of Photo-Optical Instrumentation Engineers (SPIE) Conference Series,
  114433Y

\bibitem[{{Fogarty} {et~al.}(2017){Fogarty}, {Pueyo}, {Mazoyer}, \&
  {N'Diaye}}]{Fogarty2017}
{Fogarty}, K., {Pueyo}, L., {Mazoyer}, J., \& {N'Diaye}, M. 2017, \aj, 154, 240

\bibitem[{{Gardner} {et~al.}(2023){Gardner}, {Mather}, {Abbott}, {Abell},
  {Abernathy}, {Abney}, {Abraham}, {Abraham}, {Abul-Huda}, {Acton}, {Adams},
  {Adams}, {Adler}, {Adriaensen}, {Aguilar}, {Ahmed}, {Ahmed}, {Ahmed},
  {Albat}, {Albert}, {Alberts}, {Aldridge}, {Allen}, {Allen}, {Altenburg},
  {Altunc}, {Alvarez}, {{\'A}lvarez-M{\'a}rquez}, {Alves de Oliveira},
  {Ambrose}, {Anandakrishnan}, {Andersen}, {Anderson}, {Anderson}, {Anderson},
  {Anderson}, {Aprea}, {Archer}, {Arenberg}, {Argyriou}, {Arribas}, {Artigau},
  {Arvai}, {Atcheson}, {Atkinson}, {Averbukh}, {Aymergen}, {Bacinski},
  {Baggett}, {Bagnasco}, {Baker}, {Balzano}, {Banks}, {Baran}, {Barker},
  {Barrett}, {Barringer}, {Barto}, {Bast}, {Baudoz}, {Baum}, {Beatty},
  {Beaulieu}, {Bechtold}, {Beck}, {Beddard}, {Beichman}, {Bellagama}, {Bely},
  {Berger}, {Bergeron}, {Bernier}, {Bertch}, {Beskow}, {Betz}, {Biagetti},
  {Birkmann}, {Bjorklund}, {Blackwood}, {Blazek}, {Blossfeld}, {Bluth},
  {Boccaletti}, {Boegner}, {Bohlin}, {Boia}, {B{\"o}ker}, {Bonaventura},
  {Bond}, {Bosley}, {Boucarut}, {Bouchet}, {Bouwman}, {Bower}, {Bowers},
  {Bowers}, {Boyce}, {Boyer}, {Boyer}, {Boyer}, {Boyer}, {Bradley}, {Brady},
  {Brandl}, {Brannen}, {Breda}, {Bremmer}, {Brennan}, {Bresnahan}, {Bright},
  {Broiles}, {Bromenschenkel}, {Brooks}, {Brooks}, {Brown}, {Brown}, {Brown},
  {Bruce}, {Bryson}, {Bujanda}, {Bullock}, {Bunker}, {Bureo}, {Burt}, {Bush},
  {Bushouse}, {Bussman}, {Cabaud}, {Cale}, {Calhoon}, {Calvani}, {Canipe},
  {Caputo}, {Cara}, {Carey}, {Case}, {Cesari}, {Cetorelli}, {Chance},
  {Chandler}, {Chaney}, {Chapman}, {Charlot}, {Chayer}, {Cheezum}, {Chen},
  {Chen}, {Cherinka}, {Chichester}, {Chilton}, {Chittiraibalan}, {Clampin},
  {Clark}, {Clark}, {Clark}, {Claybrooks}, {Cleveland}, {Cohen}, {Cohen},
  {Col{\'o}n}, {Coleman}, {Colina}, {Comber}, {Comeau}, {Comer}, {Conde Reis},
  {Connolly}, {Conroy}, {Contos}, {Contreras}, {Cook}, {Cooper}, {Cooper},
  {Correia}, {Correnti}, {Cossou}, {Costanza}, {Coulais}, {Cox}, {Coyle},
  {Cracraft}, {Crew}, {Curtis}, {Cusveller}, {Da Costa Maciel}, {Dailey},
  {Daugeron}, {Davidson}, {Davies}, {Davis}, {Davis}, {Day}, {de Chambure}, {de
  Jong}, {De Marchi}, {Dean}, {Decker}, {Delisa}, {Dell}, {Dellagatta},
  {Dembinska}, {Demosthenes}, {Dencheva}, {Deneu}, {DePriest}, {Deschenes},
  {Dethienne}, {Detre}, {Diaz}, {Dicken}, {DiFelice}, {Dillman}, {Disharoon},
  {Dixon}, {Doggett}, {Dominguez}, {Donaldson}, {Doria-Warner}, {Santos},
  {Doty}, {Douglas}, {Doyon}, {Dressler}, {Driggers}, {Driggers}, {Dunn},
  {DuPrie}, {Dupuis}, {Durning}, {Dutta}, {Earl}, {Eccleston}, {Ecobichon},
  {Egami}, {Ehrenwinkler}, {Eisenhamer}, {Eisenhower}, {Eisenstein}, {El
  Hamel}, {Elie}, {Elliott}, {Elliott}, {Engesser}, {Espinoza}, {Etienne},
  {Etxaluze}, {Evans}, {Fabreguettes}, {Falcolini}, {Falini}, {Fatig},
  {Feeney}, {Feinberg}, {Fels}, {Ferdous}, {Ferguson}, {Ferrarese}, {Ferreira},
  {Ferruit}, {Ferry}, {Filippazzo}, {Firre}, {Fix}, {Flagey}, {Flanagan},
  {Fleming}, {Florian}, {Flynn}, {Foiadelli}, {Fontaine}, {Fontanella},
  {Forshay}, {Fortner}, {Fox}, {Framarini}, {Francisco}, {Franck}, {Franx},
  {Franz}, {Friedman}, {Friend}, {Frost}, {Fu}, {Fullerton}, {Gaillard},
  {Galkin}, {Gallagher}, {Galyer}, {Garc{\'\i}a Mar{\'\i}n}, {Gardner},
  {Garland}, {Garrett}, {Gasman}, {G{\'a}sp{\'a}r}, {Gastaud}, {Gaudreau},
  {Gauthier}, {Geers}, {Geithner}, {Gennaro}, {Gerber}, {Gereau}, {Giampaoli},
  {Giardino}, {Gibbons}, {Gilbert}, {Gilman}, {Girard}, {Giuliano}, {Gkountis},
  {Glasse}, {Glassmire}, {Glauser}, {Glazer}, {Goldberg}, {Golimowski},
  {Gonzaga}, {Gordon}, {Gordon}, {Goudfrooij}, {Gough}, {Graham}, {Grau},
  {Green}, {Greene}, {Greene}, {Greenfield}, {Greenhouse}, {Greve}, {Greville},
  {Grimaldi}, {Groe}, {Groebner}, {Grumm}, {Grundy}, {G{\"u}del}, {Guillard},
  {Guldalian}, {Gunn}, {Gurule}, {Gutman}, {Guy}, {Guyot}, {Hack}, {Haderlein},
  {Hagan}, {Hagedorn}, {Hainline}, {Haley}, {Hami}, {Hamilton}, {Hammann},
  {Hammel}, {Hanley}, {Hansen}, {Hardy}, {Harnisch}, {Harr}, {Harris}, {Hart},
  {Hartig}, {Hasan}, {Hashim}, {Hashimoto}, {Haskins}, {Hawkins}, {Hayden},
  {Hayden}, {Healy}, {Hecht}, {Heeg}, {Hejal}, {Helm}, {Hengemihle}, {Henning},
  {Henry}, {Henry}, {Henshaw}, {Hernandez}, {Herrington}, {Heske}, {Hesman},
  {Hickey}, {Hilbert}, {Hines}, {Hinz}, {Hirsch}, {Hitcho}, {Hodapp}, {Hodge},
  {Hoffman}, {Holfeltz}, {Holler}, {Hoppa}, {Horner}, {Howard}, {Howard},
  {Huber}, {Hunkeler}, {Hunter}, {Hunter}, {Hurd}, {Hurst}, {Hutchings},
  {Hylan}, {Ignat}, {Illingworth}, {Irish}, {Isaacs}, {Jackson}, {Jaffe},
  {Jahic}, {Jahromi}, {Jakobsen}, {James}, {James}, {James}, {Jamieson},
  {Jandra}, {Jayawardhana}, {Jedrzejewski}, {Jeffers}, {Jensen}, {Joanne},
  {Johns}, {Johnson}, {Johnson}, {Johnson}, {Johnson}, {Johnson}, {Johnson},
  {Johnstone}, {Jollet}, {Jones}, {Jones}, {Jones}, {Jones}, {Jones}, {Jordan},
  {Jordan}, {Jue}, {Jurkowski}, {Justis}, {Justtanont}, {Kaleida}, {Kalirai},
  {Kalmanson}, {Kaltenegger}, {Kammerer}, {Kan}, {Kanarek}, {Kao}, {Karakla},
  {Karl}, {Kassin}, {Kauffman}, {Kavanagh}, {Kelley}, {Kelly}, {Kendrew},
  {Kennedy}, {Kenny}, {Keski-Kuha}, {Keyes}, {Khan}, {Kidwell}, {Kimble},
  {King}, {King}, {Kinzel}, {Kirk}, {Kirkpatrick}, {Klaassen}, {Klingemann},
  {Klintworth}, {Knapp}, {Knight}, {Knollenberg}, {Knutsen}, {Koehler},
  {Koekemoer}, {Kofler}, {Kontson}, {Kovacs}, {Kozhurina-Platais}, {Krause},
  {Kriss}, {Krist}, {Kristoffersen}, {Krogel}, {Krueger}, {Kulp}, {Kumari},
  {Kwan}, {Kyprianou}, {Labador}, {Labiano}, {Lafreni{\`e}re}, {Lagage},
  {Laidler}, {Laine}, {Laird}, {Lajoie}, {Lallo}, {Lam}, {LaMassa}, {Lambros},
  {Lampenfield}, {Lander}, {Langston}, {Larson}, {Larson}, {LaVerghetta},
  {Law}, {Lawrence}, {Lee}, {Lee}, {Lee}, {Leisenring}, {Leveille}, {Levenson},
  {Levi}, {Levine}, {Lewis}, {Lewis}, {Lewis}, {Libralato}, {Lidon},
  {Liebrecht}, {Lightsey}, {Lilly}, {Lim}, {Lim}, {Ling}, {Link}, {Link},
  {Lipinski}, {Liu}, {Lo}, {Lobmeyer}, {Logue}, {Long}, {Long}, {Long}, {Long},
  {L{\'o}pez-Caniego}, {Lotz}, {Love-Pruitt}, {Lubskiy}, {Luers}, {Luetgens},
  {Luevano}, {Lui}, {Lund}, {Lundquist}, {Lunine}, {L{\"u}tzgendorf}, {Lynch},
  {MacDonald}, {MacDonald}, {Macias}, {Macklis}, {Maghami}, {Maharaja},
  {Maiolino}, {Makrygiannis}, {Malla}, {Malumuth}, {Manjavacas}, {Marini},
  {Marrione}, {Marston}, {Martel}, {Martin}, {Martin}, {Martinez}, {Maschmann},
  {Masci}, {Masetti}, {Maszkiewicz}, {Matthews}, {Matuskey}, {McBrayer},
  {McCarthy}, {McCaughrean}, {McClare}, {McClare}, {McCloskey}, {McClurg},
  {McCoy}, {McElwain}, {McGregor}, {McGuffey}, {McKay}, {McKenzie}, {McLean},
  {McMaster}, {McNeil}, {De Meester}, {Mehalick}, {Meixner}, {Mel{\'e}ndez},
  {Menzel}, {Menzel}, {Merz}, {Mesterharm}, {Meyer}, {Meyett}, {Meza},
  {Midwinter}, {Milam}, {Miller}, {Miller}, {Miskey}, {Misselt}, {Mitchell},
  {Mohan}, {Montoya}, {Moran}, {Morishita}, {Moro-Mart{\'\i}n}, {Morrison},
  {Morrison}, {Morse}, {Moschos}, {Moseley}, {Mosier}, {Mosner}, {Mountain},
  {Muckenthaler}, {Mueller}, {Mueller}, {Muhiem}, {M{\"u}hlmann}, {Mullally},
  {Mullen}, {Munger}, {Murphy}, {Murray}, {Muzerolle}, {Mycroft}, {Myers},
  {Myers}, {Myers}, {Myers}, {Myrick}, {Nagle}, {Nayak}, {Naylor}, {Neff},
  {Nelan}, {Nella}, {Nguyen}, {Nguyen}, {Nickson}, {Nidhiry}, {Niedner},
  {Nieto-Santisteban}, {Nikolov}, {Nishisaka}, {Noriega-Crespo}, {Nota},
  {O'Mara}, {Oboryshko}, {O'Brien}, {Ochs}, {Offenberg}, {Ogle}, {Ohl},
  {Olmsted}, {Osborne}, {O'Shaughnessy}, {{\"O}stlin}, {O'Sullivan}, {Otor},
  {Ottens}, {Ouellette}, {Outlaw}, {Owens}, {Pacifici}, {Page}, {Paranilam},
  {Park}, {Parrish}, {Paschal}, {Patapis}, {Patel}, {Patrick}, {Pattishall},
  {Paul}, {Paul}, {Pauly}, {Pavlovsky}, {Pe{\~n}a-Guerrero}, {Pedder}, {Peek},
  {Pelham}, {Penanen}, {Perriello}, {Perrin}, {Perrine}, {Perrygo}, {Peslier},
  {Petach}, {Peterson}, {Pfarr}, {Pierson}, {Pietraszkiewicz}, {Pilchen},
  {Pipher}, {Pirzkal}, {Pitman}, {Player}, {Plesha}, {Plitzke}, {Pohner},
  {Poletis}, {Pollizzi}, {Polster}, {Pontius}, {Pontoppidan}, {Porges},
  {Potter}, {Prescott}, {Proffitt}, {Pueyo}, {Quispe Neira}, {Radich}, {Rager},
  {Rameau}, {Ramey}, {Ramos Alarcon}, {Rampini}, {Rapp}, {Rashford},
  {Rauscher}, {Ravindranath}, {Rawle}, {Rawlings}, {Ray}, {Regan}, {Rehm},
  {Rehm}, {Reid}, {Reis}, {Renk}, {Reoch}, {Ressler}, {Rest}, {Reynolds},
  {Richon}, {Richon}, {Ridgaway}, {Riedel}, {Rieke}, {Rieke}, {Rifelli},
  {Rigby}, {Riggs}, {Ringel}, {Ritchie}, {Rix}, {Robberto}, {Robinson},
  {Robinson}, {Robinson}, {Rock}, {Rodriguez}, {Rodr{\'\i}guez del Pino},
  {Roellig}, {Rohrbach}, {Roman}, {Romelfanger}, {Romo}, {Rosales}, {Rose},
  {Roteliuk}, {Roth}, {Rothwell}, {Rouzaud}, {Rowe}, {Rowlands}, {Roy},
  {Royer}, {Rui}, {Rumler}, {Rumpl}, {Russ}, {Ryan}, {Ryan}, {Saad}, {Sabata},
  {Sabatino}, {Sabbi}, {Sabelhaus}, {Sabia}, {Sahu}, {Saif}, {Salvignol},
  {Samara-Ratna}, {Samuelson}, {Sanders}, {Sappington}, {Sargent}, {Sauer},
  {Savadkin}, {Sawicki}, {Schappell}, {Scheffer}, {Scheithauer}, {Scherer},
  {Schiff}, {Schlawin}, {Schmeitzky}, {Schmitz}, {Schmude}, {Schneider},
  {Schreiber}, {Schroeven-Deceuninck}, {Schultz}, {Schwab}, {Schwartz},
  {Scoccimarro}, {Scott}, {Scott}, {Seaton}, {Seely}, {Seery}, {Seidleck},
  {Sembach}, {Shanahan}, {Shaughnessy}, {Shaw}, {Shay}, {Sheehan}, {Sheth},
  {Shih}, {Shivaei}, {Siegel}, {Sienkiewicz}, {Simmons}, {Simon}, {Sirianni},
  {Sivaramakrishnan}, {Slade}, {Sloan}, {Slocum}, {Slowinski}, {Smith},
  {Smith}, {Smith}, {Smith}, {Smith}, {Smith}, {Smolik}, {Soderblom}, {Sohn},
  {Sokol}, {Sonneborn}, {Sontag}, {Sooy}, {Soummer}, {Southwood}, {Spain},
  {Sparmo}, {Speer}, {Spencer}, {Sprofera}, {Stallcup}, {Stanley},
  {Stansberry}, {Stark}, {Starr}, {Stassi}, {Steck}, {Steeley}, {Stephens},
  {Stephenson}, {Stewart}, {Stiavelli}, {}, {Strada}, {Straughn}, {Streetman},
  {Strickland}, {Strobele}, {Stuhlinger}, {Stys}, {Such}, {Sukhatme},
  {Sullivan}, {Sullivan}, {Sumner}, {Sun}, {Sunnquist}, {Swade}, {Swam},
  {Swenton}, {Swoish}, {Tam Litten}, {Tamas}, {Tao}, {Taylor}, {Taylor}, {te
  Plate}, {Van Tea}, {Teague}, {Telfer}, {Temim}, {Texter}, {Thatte},
  {Thompson}, {Thompson}, {Thomson}, {Thronson}, {Tierney}, {Tikkanen},
  {Tinnin}, {Tippet}, {Todd}, {Tran}, {Trauger}, {Trejo}, {Vinh Truong},
  {Tsukamoto}, {Tufail}, {Tumlinson}, {Tustain}, {Tyra}, {Ubeda}, {Underwood},
  {Uzzo}, {Vaclavik}, {Valenduc}, {Valenti}, {Van Campen}, {van de Wetering},
  {Van Der Marel}, {van Haarlem}, {Vandenbussche}, {van Dishoeck},
  {Vanterpool}, {Vernoy}, {Vila Costas}, {Volk}, {Voorzaat}, {Voyton}, {Vydra},
  {Waddy}, {Waelkens}, {Wahlgren}, {Walker}, {Wander}, {Warfield}, {Warner},
  {Wasiak}, {Wasiak}, {Wehner}, {Weiler}, {Weilert}, {Weiss}, {Wells}, {Welty},
  {Wheate}, {Wheeler}, {White}, {Whitehouse}, {Whiteleather}, {Whitman},
  {Williams}, {Willmer}, {Willott}, {Willoughby}, {Wilson}, {Wilson}, {Wilson},
  {Windhorst}, {Wislowski}, {Wolfe}, {Wolfe}, {Wolff}, {Wondel}, {Woo},
  {Woods}, {Worden}, {Workman}, {Wright}, {Wu}, {Wu}, {Wun}, {Wymer},
  {Yadetie}, {Yan}, {Yang}, {Yates}, {Yeager}, {Yerger}, {Young}, {Young},
  {Yu}, {Yu}, {Zak}, {Zeidler}, {Zepp}, {Zhou}, {Zincke}, {Zonak}, \&
  {Zondag}}]{Gardner2023}
{Gardner}, J.~P., {Mather}, J.~C., {Abbott}, R., {et~al.} 2023, \pasp, 135,
  068001

\bibitem[{{Gaudi} {et~al.}(2021){Gaudi}, {Roberge}, {Habex Team}, \& {Luvoir
  Study Team}}]{Gaudi2021}
{Gaudi}, B., {Roberge}, A., {Habex Team}, \& {Luvoir Study Team}. 2021, in
  Bulletin of the American Astronomical Society, Vol.~53, 0301

\bibitem[{{Gaudi} {et~al.}(2020){Gaudi}, {Seager}, {Mennesson}, {Kiessling},
  {Warfield}, {Cahoy}, {Clarke}, {Domagal-Goldman}, {Feinberg}, {Guyon},
  {Kasdin}, {Mawet}, {Plavchan}, {Robinson}, {Rogers}, {Scowen}, {Somerville},
  {Stapelfeldt}, {Stark}, {Stern}, {Turnbull}, {Amini}, {Kuan}, {Martin},
  {Morgan}, {Redding}, {Stahl}, {Webb}, {Alvarez-Salazar}, {Arnold}, {Arya},
  {Balasubramanian}, {Baysinger}, {Bell}, {Below}, {Benson}, {Blais}, {Booth},
  {Bourgeois}, {Bradford}, {Brewer}, {Brooks}, {Cady}, {Caldwell}, {Calvet},
  {Carr}, {Chan}, {Cormarkovic}, {Coste}, {Cox}, {Danner}, {Davis}, {Dewell},
  {Dorsett}, {Dunn}, {East}, {Effinger}, {Eng}, {Freebury}, {Garcia}, {Gaskin},
  {Greene}, {Hennessy}, {Hilgemann}, {Hood}, {Holota}, {Howe}, {Huang}, {Hull},
  {Hunt}, {Hurd}, {Johnson}, {Kissil}, {Knight}, {Kolenz}, {Kraus}, {Krist},
  {Li}, {Lisman}, {Mandic}, {Mann}, {Marchen}, {Marrese-Reading}, {McCready},
  {McGown}, {Missun}, {Miyaguchi}, {Moore}, {Nemati}, {Nikzad}, {Nissen},
  {Novicki}, {Perrine}, {Pineda}, {Polanco}, {Putnam}, {Qureshi}, {Richards},
  {Eldorado Riggs}, {Rodgers}, {Rud}, {Saini}, {Scalisi}, {Scharf}, {Schulz},
  {Serabyn}, {Sigrist}, {Sikkia}, {Singleton}, {Shaklan}, {Smith}, {Southerd},
  {Stahl}, {Steeves}, {Sturges}, {Sullivan}, {Tang}, {Taras}, {Tesch},
  {Therrell}, {Tseng}, {Valente}, {Van Buren}, {Villalvazo}, {Warwick}, {Webb},
  {Westerhoff}, {Wofford}, {Wu}, {Woo}, {Wood}, {Ziemer}, {Arney}, {Anderson},
  {Ma{\'\i}z-Apell{\'a}niz}, {Bartlett}, {Belikov}, {Bendek}, {Cenko},
  {Douglas}, {Dulz}, {Evans}, {Faramaz}, {Feng}, {Ferguson}, {Follette},
  {Ford}, {Garc{\'\i}a}, {Geha}, {Gelino}, {G{\"o}tberg}, {Hildebrandt}, {Hu},
  {Jahnke}, {Kennedy}, {Kreidberg}, {Isella}, {Lopez}, {Marchis}, {Macri},
  {Marley}, {Matzko}, {Mazoyer}, {McCandliss}, {Meshkat}, {Mordasini},
  {Morris}, {Nielsen}, {Newman}, {Petigura}, {Postman}, {Reines}, {Roberge},
  {Roederer}, {Ruane}, {Schwieterman}, {Sirbu}, {Spalding}, {Teplitz},
  {Tumlinson}, {Turner}, {Werk}, {Wofford}, {Wyatt}, {Young}, \&
  {Zellem}}]{Gaudi2020}
{Gaudi}, B.~S., {Seager}, S., {Mennesson}, B., {et~al.} 2020, arXiv e-prints,
  arXiv:2001.06683

\bibitem[{{Hours} {et~al.}(2022){Hours}, {Carlotti}, {Mouillet},
  {Delboulb{\'e}}, {Guieu}, {Jocou}, {Moulin}, {Pancher}, {Rabou}, {Choquet},
  {Dohlen}, {Sauvage}, \& {N'Diaye}}]{Hours2022}
{Hours}, A., {Carlotti}, A., {Mouillet}, D., {et~al.} 2022, in Society of
  Photo-Optical Instrumentation Engineers (SPIE) Conference Series, Vol. 12185,
  Adaptive Optics Systems VIII, ed. L.~{Schreiber}, D.~{Schmidt}, \&
  E.~{Vernet}, 121852E

\bibitem[{{Jovanovic} {et~al.}(2018){Jovanovic}, {Absil}, {Baudoz}, {Beaulieu},
  {Bottom}, {Cady}, {Carlomagno}, {Carlotti}, {Doelman}, {Fogarty}, {Galicher},
  {Guyon}, {Haffert}, {Huby}, {Jewell}, {Keller}, {Kenworthy}, {Knight},
  {K{\"u}hn}, {Miller}, {Mazoyer}, {N'Diaye}, {Por}, {Pueyo}, {Riggs}, {Ruane},
  {Sirbu}, {Snik}, {Wallace}, {Wilby}, \& {Ygouf}}]{Jovanovic2018}
{Jovanovic}, N., {Absil}, O., {Baudoz}, P., {et~al.} 2018, in Society of
  Photo-Optical Instrumentation Engineers (SPIE) Conference Series, Vol. 10703,
  Adaptive Optics Systems VI, ed. L.~M. {Close}, L.~{Schreiber}, \&
  D.~{Schmidt}, 107031U

\bibitem[{{Kasdin} {et~al.}(2003){Kasdin}, {Vanderbei}, {Spergel}, \&
  {Littman}}]{Kasdin2003}
{Kasdin}, N.~J., {Vanderbei}, R.~J., {Spergel}, D.~N., \& {Littman}, M.~G.
  2003, \apj, 582, 1147

\bibitem[{{K{\"u}hn} {et~al.}(2018){K{\"u}hn}, {Patapis}, {Lu}, \&
  {Arikan}}]{Kühn2018}
{K{\"u}hn}, J., {Patapis}, P., {Lu}, X., \& {Arikan}, M. 2018, in Society of
  Photo-Optical Instrumentation Engineers (SPIE) Conference Series, Vol. 10706,
  Advances in Optical and Mechanical Technologies for Telescopes and
  Instrumentation III, ed. R.~{Navarro} \& R.~{Geyl}, 107062N

\bibitem[{{K{\"u}hn} \& {Patapis}(2022)}]{Kühn2022}
{K{\"u}hn}, J.~G. \& {Patapis}, P. 2022, \ao, 61, 9000

\bibitem[{{Laginja} {et~al.}(2022){Laginja}, {Robles}, {Barjot}, {Leboulleux},
  {Jensen-Clem}, {Brooks}, \& {Moriarty}}]{Laginja2022}
{Laginja}, I., {Robles}, P., {Barjot}, K., {et~al.} 2022, in Society of
  Photo-Optical Instrumentation Engineers (SPIE) Conference Series, Vol. 12185,
  Adaptive Optics Systems VIII, ed. L.~{Schreiber}, D.~{Schmidt}, \&
  E.~{Vernet}, 121853A

\bibitem[{{Laginja} {et~al.}(2021){Laginja}, {Soummer}, {Mugnier}, {Pueyo},
  {Sauvage}, {Leboulleux}, {Coyle}, \& {Knight}}]{Laginja2021}
{Laginja}, I., {Soummer}, R., {Mugnier}, L.~M., {et~al.} 2021, Journal of
  Astronomical Telescopes, Instruments, and Systems, 7, 015004

\bibitem[{{Leboulleux} {et~al.}(2022{\natexlab{a}}){Leboulleux}, {Carlotti},
  {Curaba}, {Delboub{\'e}}, {Jocou}, {Moulin}, {Gluck}, \&
  {Sztefek}}]{Leboulleux2022a}
{Leboulleux}, L., {Carlotti}, A., {Curaba}, S., {et~al.} 2022{\natexlab{a}}, in
  Society of Photo-Optical Instrumentation Engineers (SPIE) Conference Series,
  Vol. 12188, Advances in Optical and Mechanical Technologies for Telescopes
  and Instrumentation, 121884J

\bibitem[{{Leboulleux} {et~al.}(2022{\natexlab{b}}){Leboulleux}, {Carlotti}, \&
  {N'Diaye}}]{Leboulleux2022c}
{Leboulleux}, L., {Carlotti}, A., \& {N'Diaye}, M. 2022{\natexlab{b}}, \aap,
  659, A143

\bibitem[{{Leboulleux} {et~al.}(2022{\natexlab{c}}){Leboulleux}, {Carlotti},
  {N'Diaye}, {Bertrou-Cantou}, {Milli}, {Pourr{\'e}}, {Cantalloube},
  {Mouillet}, \& {V{\'e}rinaud}}]{Leboulleux2022b}
{Leboulleux}, L., {Carlotti}, A., {N'Diaye}, M., {et~al.} 2022{\natexlab{c}},
  \aap, 666, A91

\bibitem[{{Leboulleux} {et~al.}(2022{\natexlab{d}}){Leboulleux}, {Carlotti},
  {N'Diaye}, {Cantalloube}, {Milli}, {Bertrou-Cantou}, {Mouillet},
  {Pourr{\'e}}, \& {V{\'e}rinaud}}]{Leboulleux2022}
{Leboulleux}, L., {Carlotti}, A., {N'Diaye}, M., {et~al.} 2022{\natexlab{d}},
  in Society of Photo-Optical Instrumentation Engineers (SPIE) Conference
  Series, Vol. 12188, Advances in Optical and Mechanical Technologies for
  Telescopes and Instrumentation, 121881S

\bibitem[{{Leboulleux} {et~al.}(2018){Leboulleux}, {Sauvage}, {Pueyo}, {Fusco},
  {Soummer}, {Mazoyer}, {Sivaramakrishnan}, {N'Diaye}, \&
  {Fauvarque}}]{Leboulleux2018}
{Leboulleux}, L., {Sauvage}, J.-F., {Pueyo}, L.~A., {et~al.} 2018, Journal of
  Astronomical Telescopes, Instruments, and Systems, 4, 035002

\bibitem[{{Martinez} {et~al.}(2023){Martinez}, {Beaulieu}, {Gouvret}, {Spang},
  \& {Marcotto}}]{Martinez2023}
{Martinez}, P., {Beaulieu}, M., {Gouvret}, C., {Spang}, A., \& {Marcotto}, A.
  2023, The Messenger, 190, 55

\bibitem[{{Mazoyer} {et~al.}(2019){Mazoyer}, {Baudoz}, {Belikov}, {Crill},
  {Fogarty}, {Galicher}, {Groff}, {Guyon}, {Juanola-Parramon}, {Kasdin},
  {Leboulleux}, {Sayson}, {Mawet}, {Prada}, {Mennesson}, {N'Diaye}, {Perrin},
  {Pueyo}, {Roberge}, {Ruane}, {Serabyn}, {Shaklan}, {Siegler}, {Sirbu},
  {Soummer}, {Stark}, {Trauger}, \& {Zimmerman}}]{Mazoyer2019}
{Mazoyer}, J., {Baudoz}, P., {Belikov}, R., {et~al.} 2019, in Bulletin of the
  American Astronomical Society, Vol.~51, 101

\bibitem[{{Mazoyer} {et~al.}(2018{\natexlab{a}}){Mazoyer}, {Pueyo}, {N'Diaye},
  {Fogarty}, {Zimmerman}, {Leboulleux}, {St. Laurent}, {Soummer}, {Shaklan}, \&
  {Norman}}]{Mazoyer2018}
{Mazoyer}, J., {Pueyo}, L., {N'Diaye}, M., {et~al.} 2018{\natexlab{a}}, \aj,
  155, 7

\bibitem[{{Mazoyer} {et~al.}(2018{\natexlab{b}}){Mazoyer}, {Pueyo}, {N'Diaye},
  {Fogarty}, {Zimmerman}, {Soummer}, {Shaklan}, \& {Norman}}]{Mazoyer2018b}
{Mazoyer}, J., {Pueyo}, L., {N'Diaye}, M., {et~al.} 2018{\natexlab{b}}, \aj,
  155, 8

\bibitem[{{Mennesson} {et~al.}(2022){Mennesson}, {Bailey}, {Zellem},
  {Hildebrandt}, {Ygouf}, {Rhodes}, {Zimmerman}, {Nemati}, {Gonzalez}, {Cady},
  {Kern}, {Koch}, {Krist}, {Heydorff}, {Luchik}, {Mok}, {Morrissey},
  {Poberezhskiy}, {Riggs}, {Shi}, {Zhao}, {Akeson}, {Armus}, {Greenbaum},
  {Ingalls}, \& {Lowrance}}]{Mennesson2022}
{Mennesson}, B., {Bailey}, V.~P., {Zellem}, R., {et~al.} 2022, in Society of
  Photo-Optical Instrumentation Engineers (SPIE) Conference Series, Vol. 12180,
  Space Telescopes and Instrumentation 2022: Optical, Infrared, and Millimeter
  Wave, ed. L.~E. {Coyle}, S.~{Matsuura}, \& M.~D. {Perrin}, 121801W

\bibitem[{{Morgan} {et~al.}(2019){Morgan}, {Douglas}, {Allan}, {Bierden},
  {Chakrabarti}, {Cook}, {Egan}, {Furesz}, {Gubner}, {Haughwout}, {Holden},
  {Mendillo}, {Ouellet}, {do Vale Pereira}, {Stein}, {Thibault}, {Wu}, {Xin},
  \& {Cahoy}}]{Morgan2019}
{Morgan}, R., {Douglas}, E., {Allan}, G.~W., {et~al.} 2019, Micromachines, 10,
  366

\bibitem[{{N'Diaye} {et~al.}(2015){N'Diaye}, {Pueyo}, \&
  {Soummer}}]{N'Diaye2015}
{N'Diaye}, M., {Pueyo}, L., \& {Soummer}, R. 2015, \apj, 799, 225

\bibitem[{{N'Diaye} {et~al.}(2016){N'Diaye}, {Soummer}, {Pueyo}, {Carlotti},
  {Stark}, \& {Perrin}}]{N'Diaye2016}
{N'Diaye}, M., {Soummer}, R., {Pueyo}, L., {et~al.} 2016, \apj, 818, 163

\bibitem[{{Nickson} {et~al.}(2022){Nickson}, {Por}, {Nguyen}, {Soummer},
  {Laginja}, {Sahoo}, {Pueyo}, {St. Laurent}, {N'Diaye}, {Zimmerman}, {Noss},
  \& {Perrin}}]{Nickson2022}
{Nickson}, B.~F., {Por}, E.~H., {Nguyen}, M.~M., {et~al.} 2022, in Society of
  Photo-Optical Instrumentation Engineers (SPIE) Conference Series, Vol. 12180,
  Space Telescopes and Instrumentation 2022: Optical, Infrared, and Millimeter
  Wave, ed. L.~E. {Coyle}, S.~{Matsuura}, \& M.~D. {Perrin}, 121805K

\bibitem[{{Potier} {et~al.}(2019){Potier}, {Baudoz}, {Galicher}, {Huby}, \&
  {Singh}}]{Potier2019}
{Potier}, A., {Baudoz}, P., {Galicher}, R., {Huby}, E., \& {Singh}, G. 2019,
  arXiv e-prints, arXiv:1910.09064

\bibitem[{{Roberge} {et~al.}(2021){Roberge}, {Fischer}, {Peterson}, {Bean},
  {Calzetti}, {Dawson}, {Dressing}, {Feinberg}, {France}, {Guyon}, {Harris},
  {Marley}, {Meadows}, {Moustakas}, {O'Meara}, {Pascucci}, {Postman}, {Pueyo},
  {Redding}, {Rigby}, {Schiminovich}, {Schmidt}, {Stapelfeldt}, {Stark}, \&
  {Tumlinson}}]{Roberge2021}
{Roberge}, A., {Fischer}, D., {Peterson}, B., {et~al.} 2021, in Bulletin of the
  American Astronomical Society, Vol.~53, 332

\bibitem[{{Ruane} {et~al.}(2018){Ruane}, {Riggs}, {Mazoyer}, {Por}, {N'Diaye},
  {Huby}, {Baudoz}, {Galicher}, {Douglas}, {Knight}, {Carlomagno}, {Fogarty},
  {Pueyo}, {Zimmerman}, {Absil}, {Beaulieu}, {Cady}, {Carlotti}, {Doelman},
  {Guyon}, {Haffert}, {Jewell}, {Jovanovic}, {Keller}, {Kenworthy}, {Kuhn},
  {Miller}, {Sirbu}, {Snik}, {Wallace}, {Wilby}, \& {Ygouf}}]{Ruane2018}
{Ruane}, G., {Riggs}, A., {Mazoyer}, J., {et~al.} 2018, in Society of
  Photo-Optical Instrumentation Engineers (SPIE) Conference Series, Vol. 10698,
  Space Telescopes and Instrumentation 2018: Optical, Infrared, and Millimeter
  Wave, ed. M.~{Lystrup}, H.~A. {MacEwen}, G.~G. {Fazio}, N.~{Batalha},
  N.~{Siegler}, \& E.~C. {Tong}, 106982S

\bibitem[{{Snik} {et~al.}(2018){Snik}, {Absil}, {Baudoz}, {Beaulieu}, {Bendek},
  {Cady}, {Carlomagno}, {Carlotti}, {Cvetojevic}, {Doelman}, {Fogarty},
  {Galicher}, {Guyon}, {Haffert}, {Huby}, {Jewell}, {Jovanovic}, {Keller},
  {Kenworthy}, {Knight}, {Kuhn}, {Mazoyer}, {Miller}, {N'Diaye}, {Norris},
  {Por}, {Pueyo}, {Riggs}, {Ruane}, {Sirbu}, {Wallace}, {Wilby}, \&
  {Ygouf}}]{Snik2018}
{Snik}, F., {Absil}, O., {Baudoz}, P., {et~al.} 2018, in Society of
  Photo-Optical Instrumentation Engineers (SPIE) Conference Series, Vol. 10706,
  Advances in Optical and Mechanical Technologies for Telescopes and
  Instrumentation III, ed. R.~{Navarro} \& R.~{Geyl}, 107062L

\bibitem[{{Soummer} {et~al.}(2006){Soummer}, {Aime}, {Ferrari},
  {Sivaramakrishnan}, {Oppenheimer}, {Makidon}, \& {Macintosh}}]{Soummer2006}
{Soummer}, R., {Aime}, C., {Ferrari}, A., {et~al.} 2006, in IAU Colloq. 200:
  Direct Imaging of Exoplanets: Science \& Techniques, ed. C.~{Aime} \&
  F.~{Vakili}, 367--372

\bibitem[{{Vaughan} {et~al.}(2023){Vaughan}, {Gebhard}, {Bott}, {Casewell},
  {Cowan}, {Doelman}, {Kenworthy}, {Mazoyer}, {Millar-Blanchaer}, {Trees},
  {Stam}, {Absil}, {Altinier}, {Baudoz}, {Belikov}, {Bidot}, {Birkby}, {Bonse},
  {Brandl}, {Carlotti}, {Choquet}, {van Dam}, {Desai}, {Fogarty}, {Fowler},
  {van Gorkom}, {Gutierrez}, {Guyon}, {Haffert}, {Herscovici-Schiller},
  {Hours}, {Juanola-Parramon}, {Kleisioti}, {K{\"o}nig}, {van Kooten},
  {Krasteva}, {Laginja}, {Landman}, {Leboulleux}, {Mouillet}, {N'Diaye}, {Por},
  {Pueyo}, \& {Snik}}]{Vaughan2023}
{Vaughan}, S.~R., {Gebhard}, T.~D., {Bott}, K., {et~al.} 2023, \mnras, 524,
  5477

\end{thebibliography}

\end{document}